\documentstyle[aps,preprint]{revtex}
\begin{document}
\title{
Skyrmions in Higher Landau Levels
}
\author{X.-G. Wu}
\address{
Department of Physics, S.U.N.Y. at Stony Brook, Stony Brook, New York 11794
}
\author{S. L. Sondhi}
\address{
Department of Physics, University of Illinois at Urbana-Champaign, Urbana,
Illinois 61801-3080
}
\maketitle
\begin{abstract}
We calculate the energies of quasiparticles with large numbers of reversed
spins (``skyrmions'') for odd integer filling factors $\nu=2k+1$, $k\ge1$.
We find, in contrast with the known result for $\nu=1$ ($k=0$), that these
quasiparticles always have higher energy than the fully polarized ones
and hence are not the low energy charged excitations, even at small Zeeman
energies. It follows that skyrmions are the relevant quasiparticles only
at $\nu = 1$, $1/3$ and $1/5$.
\end{abstract}
\narrowtext
Two advances in device fabrication have, in recent years, focussed the
attention of theorists and experimentalists on the role of the spin
degree of freedom in the physics of systems that exhibit the
quantum Hall effect (QHE). These are the advent of higher mobility samples
that exhibit the QHE at relatively ``low'' magnetic fields and correspondingly
small values of the Zeeman energy, and that of double layer systems where
the layer index defines an approximately $SU(2)$ symmetric ``pseudospin''
that does not couple to a Zeeman term at all. The description of the
physics of the spin was given a new twist in recent work \cite{sondhi1} on
ferromagnetic filling factors in the QHE \cite{fn1} in which it was
argued that the low energy, long wavelength physics was distinguished
by a novel linkage between the density fluctuations and those of
the {\em topological} density of the spins; this linkage has been extended
to double layer systems and exploited to great effect by Yang {\em et al.}
\cite{yang} in their theory of a novel phase transition in the latter
\cite{murphy}.

Our focus in this communication is on the  specific use of the
density--topological density relation by Sondhi {\em et al.}, namely
their demonstration that at $\nu=1$, $1/3$ and $1/5$, and in the small
Zeeman energy ($g \rightarrow 0 $) limit, the quasiparticles involve a
large number of reversed spins and are skyrmions.
While their specific work was restricted to these filling factors and is
corroborated both by their finite size studies and by subsequent Hartree-Fock
calculations for $\nu=1$ \cite{fertig}, it was suggested
in \cite{sondhi1} that the conclusions would generalize to all
ferromagnetic fillings. It was subsequently pointed out by Jain and Wu
\cite{wu1} that finite size studies at $\nu=3$ yield fully polarized
quasiparticles and that this would continue to be the case at all higher
odd integer filling factors as well. Our purpose here is to argue that
1) there are skyrmions in the quasiparticle spectrum at odd integer
$\nu \ge 3$, but that 2) they have, in agreement with \cite{wu1}, higher
energies than the fully polarized quasiparticles and hence are not the
relevant low energy excitations in any regime. Combining this
information with general consequences of the composite fermion description
of the QHE, we conclude that the only fillings at which skyrmions are
the lowest energy excitations at small Zeeman energies are $\nu=1$, $1/3$
and $1/5$.

\noindent
{\bf Skyrmions Energies from Effective Lagrangian:} We briefly
recapitulate the procedure used to calculate the energies of the
skyrmions in \cite{sondhi1}. We begin by noting that
the long wavelength physics, at ferromagnetic fillings and at small
Zeeman energies, can be described by a purely magnetic effective lagrangian
of the form,
\begin{eqnarray}
{\cal L_{\rm eff}}= \alpha {\bf \cal A}({\bf n}({\bf r})) \cdot
\partial_{t}{\bf n}({\bf r}) +
\alpha' (\nabla {\bf n}({\bf r}))^2 + g \overline{\rho} \mu_B {\bf n}({\bf r})
 \cdot {\bf B} \nonumber  \\
 - \frac{1}{2}\int d^{2}r'\: V({\bf r}-{\bf r}') q({\bf r}) q({\bf r}') .
\label{lagrangian}
\end{eqnarray}
Here ${\bf n}({\bf r})$ is a field of unit magnitude and describes the
orientation of the local spins, ${\bf \cal A}$ is the vector potential of
a unit monopole, i.e. $\epsilon^{ ijk}\partial_{j}{\cal A}^{k}=n^i$,
and $q({\bf r})= \epsilon^{ij}\epsilon^{abc} n^{a}
\partial_{i} n^{b} \partial_{j} n^{c}/8 \pi$ is the topological density
of the ${\bf n}$ field.
The first
three terms are present in any ferromagnet; however, the last term is
specific to the quantum Hall problem and arises from the relation,
\cite{sondhi1}
\begin{equation}
\delta \rho({\bf r})= \nu q({\bf r}) \ ,
\label{dens-top}
\end{equation}
between the fluctuations of the density and those of the topological
density of the spins. The value of $\nu$ here is the filling factor
of the spins that constitute the magnetic degrees of freedom and is
therefore equal to 1 for all odd integer $\nu$.

Skyrmions (antiskyrmions) are topologically nontrivial excitations of
the spin field that carry topological charge $Q \equiv \int d^{2}r q({\bf r})
= \pm 1$. To compute their energies in the small Zeeman limit we need
to fix the values of $\alpha$ and $\alpha'$ which can be done from a
knowledge of Larmor's theorem and the spin-wave dispersion. More precisely,
if $\hbar \omega(q) \sim g \mu_B B + \kappa (e^2/\epsilon l) (ql)^2$
at small $q$, then
\begin{equation}
\alpha= \frac{1}{4} \hbar \rho \ \ {\rm and} \ \ \  \alpha'=
\frac{\kappa}{8 \pi} \frac{e^2}{\epsilon l} \ \ \ ,
\end{equation}
where $l$ is the magnetic length and $\rho$ is the density of electrons
with free spins; at odd integer $\nu$ this equals $1/(2 \pi l^2)$.
Finally, the limiting value of the energy of the skyrmions at small
Zeeman energies is $E_s = 8 \pi \alpha' = \kappa (e^2/\epsilon l)$.

We calculate the spin-wave stiffness $\kappa$ using the results of
Kallin and Halperin \cite{kalhalp}. Their Eq.\ (4.11) yields, for
$\nu = 2k+1$, the dispersion relation
\begin{equation} \label{dispersion}
\hbar \omega(q) = g \mu_B B - \Sigma_k - \tilde{V}^{(1)}_{kkkk} (q)
\end{equation}
where $\Sigma_k$ is given by the expressions \cite{sondhi1},
\begin{eqnarray} \label{selfenergy}
\Sigma_k &=& - \frac{e^2}{\epsilon l} V(k,k) \nonumber \\
V(l,m) &=& \frac{1}{\sqrt{2}\,m!} \sum_{r=0}^{l} {l \choose r}
(-1)^{r} \frac{\Gamma
(r + 1/2) \Gamma(m-r + \frac{1}{2})}{r! \,\Gamma(\frac{1}{2} -r)}
\end{eqnarray}
and the matrix element is defined as the integral
\begin{equation} \label{matrixelt}
\tilde{V}^{(1)}_{kkkk} (q) = \frac{e^2}{\epsilon l}
\int \frac{d^2x}{2 \pi} \frac{1}{|{\bf x} - {\bf q}l |}
[L_k(x^2/2)]^2 e^{-x^2/2}.
\end{equation}
Two features of $\tilde{V}^{(1)}_{kkkk}(q)$ follow from these
expressions at once: first that $\tilde{V}^{(1)}_{kkkk}(0) = \Sigma_k$
in order that the Larmor theorem holds, and second that
$\tilde{V}^{(1)}_{kkkk}(q)$ vanishes as $q \rightarrow \infty$ whence
the gap to creating a polarized quasiparticle-quasihole pair is exactly
$-\Sigma_k$.

We have not carried out the integral in (\ref{matrixelt}) as it stands.
Instead we expand the Laguerre polynomials as
\begin{eqnarray}
L_k(x) &=& \sum_{m=0}^k (-1)^m {k \choose m} \frac{x^m}{m!} \nonumber \\
       &\equiv & \sum_{m=0}^k (-1)^m c^k_m x^m
\end{eqnarray}
and find that
\begin{equation}
\tilde{V}^{(1)}_{kkkk}(q) = \sqrt{\frac{\pi}{2}}
\sum_l \sum_m c^k_l c^k_m
\frac{d^{l+m}}{d \tau^{l+m}} \frac{1}{\sqrt{\tau}}
e^{-\tau q^2 l^2/4} I_o(\alpha q^2 l^2/4) |_{\tau=1}.
\end{equation}
This is evidently an expansion in powers of $(q l)^2$ and using
{\em Mathematica} it is straightforward to  obtain the coefficient of
the quadratic term for a given filling factor. For the first few odd
integer fillings we find $\kappa = {1\over 4} \sqrt{\pi \over 2}$ ($\nu=1$),
$ {7\over 16} \sqrt{\pi\over 2}$ ($\nu= 3$), $ {145\over 256}
\sqrt{\pi\over 2}$ ($\nu=5$). These are also, in units of $e^2/\epsilon l$,
the energies of the skyrmions/antiskyrmions and therefore {\em half} the gap
to making an (infinite) skyrmion-antiskyrmion pair.

For the corresponding fillings we obtain the gap to a fully
polarized quasiparticle-quasihole pair from Eq.\ (\ref{selfenergy}); these are,
 in units of $e^2/\epsilon l$, $\sqrt{\pi \over 2}$ ($\nu=1$),
${3 \over 4} \sqrt{\pi \over 2}$ ($\nu=3$),
${41 \over 64}  \sqrt{\pi \over 2}$ ($\nu=5$). Finally, we find that
the ratios of the interaction energy of an (infinite) skyrmion-antiskyrmion
pair to that of a fully polarized quasiparticle-quasihole pair are:
\begin{equation}
{\Delta_{\rm sk-ask} \over \Delta_{\rm pqh-pqe}} =
{1 \over 2} \ (\nu =1), \ \  {7 \over 6} \ (\nu=3) \ \  {\rm and} \ \
{145 \over 82} \ (\nu=5) \ .
\end{equation}
It follows then, that of the odd integer filling factors only $\nu=1$
has large skyrmions as its lowest energy quasiparticles at small Zeeman
energies.

\noindent
{\bf Finite Size Study at $\nu=3$:} For the particular case of $\nu=3$,
we have expanded on the $g=0$ finite size work of Jain and Wu on the sphere
by studying systems with sizes ranging from 4 to 10 particles for the
quasihole and 6 to 12 particles for the quasielectron. Here we assume that
the lowest Landau level states of both spins are filled and inert and hence
we restrict ourselves to the states in the second Landau level alone.
We confirm that the ground states in the quasihole and quasielectron
sectors, i.e. $\pm 1$ flux quantum away from commensuration ($\nu =3$),
are maximally spin polarized consistent with the restricted Hilbert space
and Fermi statistics.
We find that in both sectors there is a low lying state  with quantum
numbers $L=0$ and $S=0$ in the spectrum. In fact these states are related,
they are particle-hole conjugates of each other; consequently, we only
describe here the results for the quasiholes (antiskyrmion) \cite{fn2}.
The candidate antiskyrmion state does in fact display spin correlations
characteristic of the infinite antiskyrmion (Fig~1). While the creation
energy of this state as well as that of the fully polarized quasihole
have considerable finite size dependence (Fig~2), the difference of their
energies is quite linear in $1/N$ except at the smallest system size (Fig~3).
For the quasihole the extrapolated values for $N=\infty$ are $0.163
e^2/\epsilon l$ for a linear fit and $0.153 e^2/\epsilon l$ for a quadratic
one which compare favorably with the analytic value of
${1 \over 8} \sqrt{\pi \over 2} (\approx 0.157) e^2/\epsilon l$. Hence even
though the
antiskyrmion is not the ground state in the quasihole sector, it is well
described by the long wavelength action. (The same is evidently true of
the skyrmion.)

\noindent
{\bf Discussion:} We have shown that at odd integer filling factors greater
than one, there are skyrmions in the spectrum of the quasiparticle sectors
as a consequence of the ferromagnetic ground states. We have
also shown that these have higher energy than the fully polarized
quasiparticles. Consequently, we do not expect them to show up in activation
energies for transport, and more generally, in the asymptotically low
temperature behavior of the system. However, the spin polarization at
finite temperatures will be quite sensitive even to a small density of
large skyrmions. For example at $\nu=3$ the energy difference between
the fully polarized quasiparticles and the skyrmions is approximately
$4 (B[T])^{1/2}$ K and so one should expect to see an anomalous decrease
of the spin polarization at temperatures of about 10 K in currently
used GaAs samples.

The reader might be somewhat puzzled that a Lagrangian that is claimed to
capture the low energy, long wavelength fluctuations does not appear to
yield the correct quasiparticles. The problem here is not that the true
quasiparticles are absent from the Landau-Ginzburg approach but that
their energies cannot be calculated reliably by means of the effective
Lagrangian.
More precisely, the polarized quasiparticles are contained in
the Landau-Ginzburg description as microscopic skyrmions \cite{fn4}.
However, the long wavelength physics that
(\ref{lagrangian}) captures correctly, is the physics of slowly varying
spin textures which only carry a small amount of local charge. To
compute accurately the energy of small skyrmions, whose density profile
varies rapidly on the scale of the magnetic length, we would have to
keep higher derivative terms and we have no practical way of doing that.
Consequently, there is no way to tell, from within the effective Lagrangian
approach, that the lowest energy quasiparticles are not slowly varying
textures.

Finally, it was already noted in \cite{sondhi1} that numerical studies
and long wavelength calculations at $\nu=1/3$ and $1/5$ indicate that
skyrmions are the relevant quasiparticles in the small Zeeman energy limit.
This conclusion also follows from the composite fermion interpretation
of the spectrum for fractional fillings \cite{jain1,wu2}. The latter
however also implies that skyrmions are {\em not} the lowest energy
quasiparticles at all other polarized fractional fillings \cite{fn3} for
they correspond
to odd integer filled Landau levels of composite fermions; a direct
numerical examination of the simplest such fraction, $\nu=3/5$, suggests
that this is indeed the case.

{\it Acknowledgments} --- We are grateful to Jainendra Jain, Anders
Karlhede, Steven Kivelson and Edward Rezayi for helpful discussions.
This work was supported in part by NSF grant Nos. DMR 93-18739 (XGW)
and DMR 91--22385 and DMR 91--57018 (SLS).



\begin{figure}
\caption{Spin (solid curve) and density (dashed curve) correlation functions
in the second Landau level antiskyrmion (quasihole).
The one-quasihole sector has, in this case, 10 electrons
and 8 flux quanta (11 available states). $\sigma_z$ is twice the spin density
and $\rho$ the density of the electrons, in units in which the filled second
Landau level has density $\rho_o = 1$. $\theta$ is the polar angle.}

\label{fig1}
\end{figure}

\begin{figure}
\caption{Polarized quasihole gaps (filled circles) and antiskyrmion gaps
(open circles) for systems with $N=$4,6,8 and 10 particles. These gaps are
the differences between the energies in the one-quasihole sector states and
the ground state with one fewer flux quantum and the same number of particles.}
\label{fig2}
\end{figure}

\begin{figure}
\caption{Difference in energy between the polarized quasihole and the
antiskyrmion for systems with $N=$4,6,8 and 10 particles.
The solid line is a linear fit to the three largest system sizes and
extrapolates to $0.163 e^2/\epsilon l$ at $N=\infty$. (A quadratic fit
yields $0.153 e^2/\epsilon l$ and gives a rough estimate of the error in
the extrapolation.) The Landau-Ginzburg analysis gives $0.157 e^2/\epsilon l$.}
\label{fig3}
\end{figure}


\begin{references}
\bibitem{fn1} A ferromagnetic filling factor is defined as one where
the ground state retains a net spin polarization even when the coupling
of the electron spin to the the external magnetic field is switched off
(i.e. $g=0$).
\bibitem{fn2} Our choice is motivated by the better scaling of the data
as a function of the number of particles $N$. On account of the the
symmetry one cannot fit both sets independently, i.e. the quasielectron data
scales equally well as a function of $N+2$.
\bibitem{fn4} There is a subtlety here. Strictly speaking, the polarized
quasihole is not a microscopic antiskyrmion but is a vortex in the parent
bosonic Chern-Simons Lagrangian \cite{sondhi1}. This is not important for
our purposes.
\bibitem{fn3} Strictly speaking this conclusion applies only at the lowest
level of the Jain hierarchy but that does cover the cases of practical
interest at this point.
\end{references}
\end{document}